%% file: main.tex
\documentclass{article}
\usepackage{spconf,amsmath,graphicx}
\usepackage{xcolor}
\usepackage{bm}
\usepackage{amsmath,graphicx}
\usepackage{amssymb}
\usepackage{multirow}
\usepackage{xcolor}
\usepackage{tabularx}
\usepackage[skip=0.5\baselineskip]{caption}
\usepackage{enumitem}
\usepackage{booktabs}
\usepackage{listings}
\usepackage{setspace} 

\usepackage{spconf,amsmath,graphicx}
\usepackage{amsmath,amsfonts,bm}
\usepackage{amsmath,graphicx}
\usepackage{amssymb}
\usepackage{multirow}
\usepackage{xcolor}
\usepackage{tabularx}
\usepackage[skip=0.5\baselineskip]{caption}
\usepackage{enumitem}
\usepackage{booktabs}
\usepackage{listings}
\usepackage{setspace} 

\newcommand{\fade}[1]{\color{lightgray} \it{#1}}

\newlength{\bibitemsep}
\newlength{\bibparskip}
\let\oldthebibliography\thebibliography
\renewcommand\thebibliography[1]{%
  \oldthebibliography{#1}%
  \setlength{\parskip}{\bibitemsep}%
  \setlength{\itemsep}{\bibparskip}%
}

\title{Best of both worlds: Multi-task Audio-Visual Automatic Speech Recognition and Active Speaker Detection}
\name{Otavio Braga\textsuperscript{1}\thanks{\textsuperscript{1}obraga@google.com}, Olivier Siohan}
\address{Google, Inc.}

\input{math_commands.tex}

\begin{document}
%\ninept
%
\maketitle
\begin{abstract}
Under noisy conditions, {\it automatic speech recognition} (ASR) can greatly benefit from the addition of visual signals coming from a video of the speaker's face. However, when multiple candidate speakers are visible this traditionally requires solving a separate problem, namely {\it active speaker detection} (ASD), which entails selecting at each moment in time which of the visible faces corresponds to the audio. Recent work has shown that we can solve both problems simultaneously by employing an attention mechanism over the competing video tracks of the speakers' faces, at the cost of sacrificing some accuracy on active speaker detection. This work closes this gap in active speaker detection accuracy by presenting a single model that can be jointly trained with a multi-task loss. By combining the two tasks during training we reduce the ASD classification accuracy by approximately $25\%$, while simultaneously improving the ASR performance when compared to the multi-person baseline trained exclusively for ASR.
\end{abstract}
\section{Introduction}
\label{sec:intro}

Complementing the acoustic signal with a video of the speaker's face is a useful strategy for ASR under noisy conditions \cite{Makino19,Afouras_2018,Chung_2017,Potamianos2003,Saenko2006,Harte2015,Serdyuk21}. In a realistic setting, however, multiple faces are potentially simultaneously on screen and one must decide which speaker face video to feed to the model. This first step has been traditionally treated as a separate problem~\cite{ChungSyncNet_2017,ChungSyncnet2_2019}, but recent work has shown that we can obtain this association with an end-to-end model employing an attention mechanism over the candidate faces~\cite{Braga20, Braga21}.

As a side effect, this attention mechanism implicitly captures the correspondence between the audio and the video of the active speaker and, thus, can also be used as an active speaker detection (ASD) model in addition to ASR. However, while interesting in itself, despite having nontrivial accuracy, the implicit association doesn't perform as well as when the attention is trained explicitly for active speaker detection (for a detailed study, see~\cite{Braga21}). ASD essentially provides a strong signal for diarization, and high ASD accuracy as a side-product is a compelling reason to include the visual signal in ASR models. Having a single model or at least sharing the visual frontend between the two tasks is particularly important since video processing in the frontend is computationally expensive.

The present work closes this gap by addressing the following question: Can we get both high accuracy on active speaker detection (ASD) and low word error rate (WER) by training a single multi-person A/V ASR model in a multi-task setting? We present a multi-task learning (MTL)~\cite{Caruana93multitasklearning} setup for a model that can simultaneously perform audio-visual ASR and active speaker detection, improving previous work on multi-person audio-visual ASR. We show that combining the two tasks is enough to significantly improve the performance of the model in the ASD task relative to the baseline in~\cite{Braga20,Braga21}, while not degrading the ASR performance of the same model trained exclusively for ASR.

\section{Model}
\label{sec:model}
\begin{figure}[!t]

\begin{minipage}[b]{1.0\linewidth}
  \centering
  \centerline{\includegraphics[width=5.0cm]{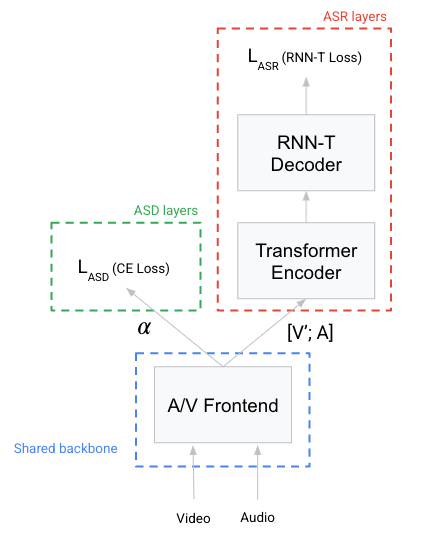}}
%  \vspace{2.0cm}
  %\centerline{(a) Multi-task model for ASR and ASD.}\medskip
\end{minipage}
\caption{{\bf Multi-task model for A/V ASR and ASD.} $\alpha \in \R^{B\times T\times M}$ is the attention tensor, indicating the probability of each of the $M$ parallel video tracks being the active speaker for each of the $B$ audio queries and each timestep. $\tV' \in \R^{B\times T \times D_v}$ is the tensor with the attention weighted visual features, and $\tA \in \R^{B\times T \times D_a}$ are the acoustic features.}
\label{fig:model}
\vspace{-4mm}
\end{figure}

Figure~\ref{fig:model} shows an overview of our model, and we explain in detail each component next.

\subsection{A/V Backbone: Shared Audio-Visual Frontend}

\noindent
{\bf Acoustic Features.} We employ log mel filterbank features as input to our trainable models. The 16kHz audio is framed with 25ms windows smoothed with the Hann window function, with strides of 10ms between frames. We compute log mel filter bank features with 80 channels, and fold every 3 consecutive feature vectors together, yielding a 240-dimensional feature vector every 30ms ($\approx 33.3$Hz). The resulting acoustic features tensor will be denoted by $\tA \in \R^{B\times T \times D_a}$, where $B$ is the batch size, $T$ is the number of time steps and $D_a$ ($=240$) the dimension of the acoustic features. During training, sequences of different lengths within a batch are zero-padded and limited to 512 steps.

\

\noindent
{\bf Audio and Video Synchronization.} Since the videos in our training set have frame rates ranging from around 23 to 30 fps, in order to make the input uniform we resample the videos in the temporal dimension to the acoustic features sample rate (33.3Hz). The resampling is performed with nearest neighbor interpolation.  In the spatial dimension, we crop the full face tracks around the mouth region to generate images of resolution $128\times128$, with RGB channels normalized between $-1$ and $1$.

\begin{table}[!t]
\small
\begin{center}
\begin{tabularx}{\linewidth}{cccccc}
\toprule
{\bf Layer} & {\bf Kernel} & {\bf Output} & {\bf Spatial} & {\bf Normalization} \\
  & {\bf Size} & {\bf Channels} & {\bf Pooling?} & {\bf Groups} \\
\midrule
0 & [1, 3, 3] & 23 & True & 1 \\
1 & [3, 1, 1] & 64 & False & 32 \\
2 & [1, 3, 3] & 64 & True & 1 \\
3 & [3, 1, 1] & 128 & False & 32 \\
4 & [1, 3, 3] & 256 & True & 1 \\
5 & [3, 1, 1] & 256 & False & 32 \\
6 & [1, 3, 3] & 921 & False & 1 \\
7 & [3, 1, 1] & 512 & False & 32 \\
8 & [1, 3, 3] & 460 & True & 1 \\
9 & [1, 1, 1] & 512 & False & 32 \\
\midrule
\bottomrule
\end{tabularx}
\caption{{\bf Configuration of (2+1)D ConvNet used to compute visual features.} Max pooling is over a $2\times2$ window on the spatial dimensions only. Additionally, we use a stride of 2 on both spatial dimensions on the first layer. For all layers, we use `VALID' paddings in the spatial dimensions, and `SAME' paddings on the temporal dimension.}
\label{table:convnet}
\end{center}
\vspace{-6mm}
\end{table}

\

\noindent
{\bf Visual Features.} For the visual frontend, we compute visual features $\tV \in \R^{M\times T\times D_v}$ with a 10-layer (2+1)D ConvNet \cite{Lecun98,Tran18} on top of the synchronized video, where $M$ is the number of competing face tracks and $D_v$, the dimension of the visual features. The exact parameters of the ConvNet can be found on Table \ref{table:convnet}. {\it This is an important deviation from~\cite{Braga21}, where blocks of 3D convolutions were used instead}. (2+1)D convolutions not only yield better performance, but are also less TPU memory intensive, allowing training with larger batch sizes, which has shown to be particularly important for obtaining lower word error rates. Note that each convolutional block is followed by group normalization~\cite{wu2018group}, and the number of groups on each layer is indicated on the table.

\

\noindent
{\bf Attention Mechanism.} Visual features of competing faces are the slices $\tV_{m,:,:} \in \R^{T\times D_v}$ along the first dimension of the visual features tensor. We employ an attention \cite{BahdanauCB14,vaswani2017attention} module in order to soft-select the one matching the audio. The attention queries $\tQ \in \R^{B\times T\times D_q}$ are computed with a 1D ConvNet of 5 layers on top of the acoustic features (consult~\cite{Braga20} for the exact parameters). We compute the attention scores with
\begin{equation}
S_{btm} = Q_{btq}W_{qv}V_{mtv}, \quad \textrm{with} \quad \tS \in \R^{B\times T\times M},
\end{equation}
in Einstein summation notation, where $\tW \in \R^{D_q \times D_v}$ is a trainable model parameter used for the bilinear function.

The attention scores are then normalized with a softmax function over the last dimension:
\begin{equation}
\label{eq:softmax}
\alpha_{btm} = \frac{e^{S_{btm}}}{\sum_{l} e^{S_{btl}}}, \quad \textrm{with} \quad \alpha \in \R^{B\times T \times M}.
\end{equation}
which are then used to yield the attention weighted visual features corresponding to each audio query with the weighted sum
\begin{equation}
V'_{btv} = \alpha_{btm} V_{mtv}, \quad \textrm{with} \quad \tV' \in \R^{B \times T \times D_v}.
\label{eq:weighted_sum}
\end{equation}

During training, we have matched pairs of audio and a single corresponding face track, so $M$ is equal to the batch size $B$. During inference, $B = 1$ and $M$ is equal to the number of parallel face tracks in the video. The visual feature that is then fed to the ASR model consists of a soft average of all the face-specific visual features, with the largest weight being attributed to the face that most likely matches the audio"

\subsection{ASR Model}

For ASR, the weighted visual features and input acoustic features are then concatenated along the last dimension, producing audio-visual features $\tF = [\tA; \tV'] \in \R^{B\times T \times (D_a + D_v)}$, which are then fed to the ASR encoder.

\

\noindent
{\bf Encoder.} For ASR we use a 14 layer Transformer encoder~\cite{vaswani2017attention}. Due to the quadratic complexity of attention over long sequences, we limit the attention context to 100 steps to the left and right of each timestep. We use 8 attention heads, each with dimension of 64, and a model dimension of 1024. This is another important improvement from~\cite{Braga21}, where a 6-layer BiLSTM encoder was used instead.

\

\noindent
{\bf Decoder.} We employ a standard RNN-T decoder \cite{graves2012sequence, graves2013speech}, with a stack of 2 LSTM layers of 2048 units and character tokens as input, with a vocabulary of size 128.

\subsection{ASD Model}

For ASD, the attention scores tensor $\tS$ is used directly for the model prediction. For each audio query and each timestep, $\tS$ gives a measure of how well each candidate video corresponds to the audio. During inference, the index of the selected video track is given by
\begin{equation}
\label{eq:one_hot}
    I_{bt} = \argmax_m S_{btm}, \quad \tI \in \R^{B\times T}.
\end{equation}

\vspace{-6mm}
\section{Multi-Task Loss for A/V ASR and ASD}
\label{sec:loss}

In this section, we describe how our model is trained by using a multi-task formulation to combine both ASR and ASD related losses.

\

\noindent
{\bf ASD.} For active speaker detection, the normalized attention weights $\alpha_{btm}$ can be used to train the attention module directly with cross entropy loss. Since during training we have pairs of corresponding audio and video from a single speaking face, we can write
\begin{equation}
\label{eq:CE}
L_{\textrm{ASD}} = \frac{1}{BT}\sum_{b=1}^B\sum_{t=1}^T\sum_{m=1}^M-[b = m]\log \alpha_{btm}.
\end{equation}

\noindent
{\bf ASR.} During training, given (audio, video, transcript) triplets, we compute the ASR loss ($L_{\textrm{ASR}}$) with the RNN-T loss~\cite{graves2012sequence, graves2013speech}. RNN-T loss computes the negative log-loss of the target transcript by computing all possible input-output alignments with a forward-backward algorithm. For the sake of brevity, we refer the reader to~\cite{graves2012sequence, graves2013speech} for details.

\

\noindent
{\bf MTL Loss.} We combine the ASD and ASR losses with a weighted linear sum of the losses for each task:
\begin{equation}
\label{eq:CE}
L = \gamma L_{\textrm{ASR}} + (1-\gamma) L_{\textrm{ASD}},
\end{equation}
where $0 \leq \gamma \leq 1$ is a blending weight between the two loss functions. $\gamma = 0$ corresponds to the model trained purely for ASD, while $\gamma = 1$, purely for ASR.

\

\noindent
{\bf Training.} Given a value of $\gamma$, we initialize the training with a checkpoint from a model trained with a single video track for A/V ASR. We use the Adam optimizer~\cite{KingmaB14} with a constant learning rate of $0.0002$ for 200k steps, and a batch size of 8 per TPU core.

\section{Datasets}
\label{sec:datasets}

\subsection{Training and Evaluation Data}

\noindent
{\bf Training.} For training, we use over 90k hours of transcribed short YouTube video segments extracted with the semi-supervised procedure originally proposed in \cite{Liao2013} and extended in \cite{Makino19,Shillingford_2019} to include video. We extract short segments where the force-aligned user uploaded transcription matches the transcriptions from a production quality ASR system. From these segments we then keep the ones in which the face tracks match the audio with high confidence. We refer the reader to~\cite{Makino19,Shillingford_2019} for more details of the pipeline.
 
\

\noindent
{\bf Evaluation.}
For the videos in both of our evaluation sets, we track the faces on screen and pick the segments with matching audio and video tracks with the same procedure used to extract the training data. Therefore, by design, the faces extracted from the video correspond with high probability to the speaker in the audio. We rely on two base datasets:
\begin{itemize}[leftmargin=*]
    \item {\it YTDEV18} \cite{Makino19}: Composed of 25 hours of manually transcribed YouTube videos, not overlapping with the training set, containing around 20k utterances.
    
    \item {\it LRS3-TED Talks} \cite{Afouras18d}: This is the largest publicly available dataset for A/V ASR, so we evaluate on it as well for completeness.
\end{itemize}

\subsection{Augmenting the Evaluation Sets with Parallel Video Tracks and Noise}
\label{sec:ms-datasets}
In order to evaluate our model in the scenario where multiple face tracks are simultaneously visible in a video, we construct a new evaluation dataset as follows: On the single track evaluation sets described in the previous section, at time $t$ both the acoustic and visual features from the corresponding face are available. To build a dataset with $N$ parallel face tracks we start from the single track set, and for every pair of matched audio and face video track we randomly pick other $N-1$ face tracks from the same dataset. Therefore, during evaluation at each time step we have the acoustic features computed from the audio and $N$ candidate visual features, without knowing which one matches the audio. We generate separate datasets for $N = 1, 2, 4, 8$ to simulate a scenario where multiple on-screen faces are competing with the face of the target speaker.

Moreover, in order to  measure the impact of the visual modality, we also evaluate on noisy conditions by adding babble noise randomly selected from the NoiseX dataset \cite{Varga1993AssessmentFA} at 0dB, 10dB and 20dB to each utterance.

\begin{table*}[!t]
\small
\setlength{\tabcolsep}{5.5pt}
\caption{Top-1 Face track selection accuracy at the frame level (ACC) and word error rate (WER) for the various noise levels, number of competing face tracks, and loss weights $\gamma$. $\gamma = 0.0$ corresponds to a model trained only with ASD loss, while $\gamma = 1.0$, only with ASR loss.}
\vspace{-4mm}
\label{table:results}
\begin{center}
\begin{tabularx}{\linewidth}{ccccccccccccccc}
\toprule
\multirow{3}{*}{\bf Dataset}    & \multirow{3}{*}{\bf Noise} & \multirow{3}{*}{\bf Tracks} & \multicolumn{10}{c}{\bf $\gamma$} & \multirow{2}{*}{\shortstack{\bf Audio \\ \bf Only}} & \multirow{2}{*}{\bf \shortstack{1-Track \\ A/V}} \\
\cmidrule{4-13}
                                &                            &                             & \multicolumn{2}{c}{0.0} & \multicolumn{2}{c}{0.25} & \multicolumn{2}{c}{0.5} & \multicolumn{2}{c}{0.75} & \multicolumn{2}{c}{1.0} & & \\
                                &                            &                             & ACC          & WER        & ACC          & WER          & ACC          & WER          & ACC          & WER         & ACC          & WER          & WER                    & WER \\
\midrule
\multirow{17}{*}{YTDEV18}	    &		                     & 1                           & \fade{1.00 } & \fade{-- } & \fade{1.00 } & 13.17        & \fade{1.00 } & 13.01        & \fade{1.00 } & 13.12       & \fade{1.00 } & {\bf 12.92 } &                        & 13.09 \\
                                &		                     & 2	                       & {\bf 0.99 }  & \fade{-- } & {\bf 0.99 }  & 13.19        & 0.98         & 13.03        & 0.97         & 13.16       & 0.89         & {\bf 12.98 } & \multirow{2}{*}{13.62} & \fade{-- }    \\
                                &		                     & 4	                       & {\bf 0.98 }  & \fade{-- } & 0.97         & 13.23        & 0.96         & {\bf 13.03 } & 0.93         & 13.19       & 0.77         & 13.15        &                        & \fade{-- }    \\
                                &		                     & 8                           & {\bf 0.95 }  & \fade{-- } & 0.94         & 13.24        & 0.92         & {\bf 13.13 } & 0.88         & 13.29       & 0.64         & 13.26        &                        & \fade{-- }    \\
\cmidrule{2-15}
                                & \multirow{4}{*}{20dB}      & 1                           & \fade{1.00 } & \fade{-- } & \fade{1.00 } & 13.30        & \fade{1.00 } & 13.23        & \fade{1.00 } & 13.30       & \fade{1.00 } & {\bf 13.14 } &                        & 13.24 \\
                                &		                     & 2	                       & {\bf 0.99 }  & \fade{-- } & 0.98         & 13.34        & 0.98         & 13.27        & 0.97         & 13.34       & 0.89         & {\bf 13.17 } & \multirow{2}{*}{13.97} & \fade{-- }    \\
                                &		                     & 4	                       & {\bf 0.97 }  & \fade{-- } & 0.96         & 13.38        & 0.95         & {\bf 13.32 } & 0.93         & 13.41       & 0.77         & 13.33        &                        & \fade{-- }    \\
                                &		                     & 8                           & {\bf 0.95 }  & \fade{-- } & 0.93         & 13.40        & 0.91         & {\bf 13.34 } & 0.87         & 13.49       & 0.64         & 13.56        &                        & \fade{-- }    \\
\cmidrule{2-15}
                                & \multirow{4}{*}{10dB}      & 1                           & \fade{1.00 } & \fade{-- } & \fade{1.00 } & 14.42        & \fade{1.00 } & 14.42        & \fade{1.00 } & 14.56       & \fade{1.00 } & {\bf 14.38 } &                        & 14.32 \\
                                &		                     & 2	                       & {\bf 0.98 }  & \fade{-- } & {\bf 0.98 }  & 14.48        & 0.97         & {\bf 14.44 } & 0.95         & 14.62       & 0.87         & 14.51        & \multirow{2}{*}{16.34} & \fade{-- }    \\
                                &		                     & 4	                       & {\bf 0.96 }  & \fade{-- } & 0.94         & 14.60        & 0.93         & {\bf 14.59 } & 0.90         & 14.72       & 0.74         & 14.93        &                        & \fade{-- }    \\
                                &		                     & 8                           & {\bf 0.92 }  & \fade{-- } & 0.90         & 14.78        & 0.87         & {\bf 14.70 } & 0.82         & 14.96       & 0.61         & 15.53        &                        & \fade{-- }    \\
\cmidrule{2-15}
                                & \multirow{4}{*}{0dB}       & 1                           & \fade{1.00}  & \fade{-- } & \fade{1.00}  & {\bf 22.58 } & 1.00         & 22.59        & \fade{1.00 } & 23.04       & \fade{1.00 } & 23.57        &                        & 21.33 \\
                                &		                     & 2	                       & {\bf 0.92 }  & \fade{-- } & 0.91         & 23.72        & 0.89         & {\bf 23.64 } & 0.87         & 24.24       & 0.79         & 25.11        & \multirow{2}{*}{36.63} & \fade{-- }    \\
                                &		                     & 4	                       & {\bf 0.84 }  & \fade{-- } & 0.81         & 25.61        & 0.79         & {\bf 25.27 } & 0.73         & 26.19       & 0.61         & 28.94        &                        & \fade{-- }    \\
                                &		                     & 8                           & {\bf 0.75 }  & \fade{-- } & 0.71         & 27.78        & 0.68         & {\bf 27.63 } & 0.61         & 28.98       & 0.47         & 34.00        &                        & \fade{-- }    \\
\cmidrule{1-15}
\multirow{4}{*}{TED.LRS3}	    &		                     & 1                           & \fade{1.00} & \fade{-- }  & \fade{1.00}  & 3.19         & \fade{1.00}  & 3.10         & \fade{1.00}  & {\bf 2.89}  & \fade{1.00}  & 2.96         &                        & 2.97          \\
                                &		                     & 2	                       & {\bf 0.99}  & \fade{-- }  & 0.98         & 3.18         & 0.97         & 3.13         & 0.97         & {\bf 2.89}  & 0.92         & 2.96         & \multirow{2}{*}{3.45}  & \fade{-- }    \\
                                & 		                     & 4	                       & {\bf 0.96}  & \fade{-- }  & {\bf 0.96}   & 3.20         & 0.93         & 3.13         & 0.93         & {\bf 2.88}  & 0.82         & 2.94         &                        & \fade{-- }    \\
                                &		                     & 8                           & {\bf 0.93}  & \fade{-- }  & 0.92         & 3.16         & 0.88         & 3.16         & 0.86         & {\bf 2.96}  & 0.70         & 3.18         &                        & \fade{-- }    \\
\bottomrule
\end{tabularx}
\end{center}
\vspace{-8mm}
\end{table*}

\section{Results}
\label{sec:results}

We train separate models with different blending weights $\gamma$ between the loss functions, and evaluate for both ASD and ASR tasks. The results are summarized on Table~\ref{table:results}. $\gamma = 0$ corresponds to a model trained purely for active speaker detection, and, thus, serves as an upper bound to the accuracy we can hope to achieve on the ASD task for the multi-task model. On the other extreme, $\gamma = 1$ corresponds to a model trained purely with ASR loss, where the A/V attention frontend provides an ASD signal, despite not being optimized for an ASD task, similar to the work in~\cite{Braga20,Braga21}.

Additionally, we include the WERs for two baseline models on the two rightmost columns of Table~\ref{table:results}: for a model trained only with audio, and for another trained with a single video track of the speaker always matching the audio. The former serves as an upper bound to the WER we need to achieve with an A/V ASR model, since once the WER surpasses the audio-only level, including the video on an ASR model becomes unjustified. 

\

\vspace{-2mm}
\noindent
{\bf Analysis.} We clearly closed the gap on ASD by combining the two losses. For $\gamma < 1.0$, there's a clear improvement on all models in terms of ASD accuracy when compared to the baseline model ($\gamma = 1$). For instance, with $\gamma = 0.5$ we see a relative improvement on average at clean, 20dB, 10dB and 0dB of approximately $26\%$, $25\%$, $26\%$ and $29\%$ in relation to the baseline multi-track model from~\cite{Braga21} (corresponding to $\gamma = 1$ on the table above). When compared to the pure ASD baseline ($\gamma = 0$), our new model shows a relative average degradation of only $2.1\%$, $2.5\%$, $3.3\%$ and $6.6\%$ in accuracy at the same noise levels, which is significantly lower than the degradations of the pure ASR model baseline ($\gamma = 1$) from ~\cite{Braga21} ($21.3\%$, $21.1\%$, $22.6\%$ and $26.2\%$ at clean, 20dB, 10dB and 0dB). Similarly, on TED.LRS3, with $\gamma = 0.75$, we see an improvement of $13.9\%$, $18.1\%$ and $22.9\%$ in ASD accuracy in relation to the baseline from~\cite{Braga21} ($\gamma = 1$), while also slightly improving the ASR WER.

For ASR, not only we do not observe a significant degradation in WER overall when combining the two losses, but, on the contrary, the auxiliary ASD loss seems to help the ASR training, actually lowering the WER in most scenarios. For instance, with $\gamma = 0.5$, we have a relative increase in accuracy by $0.2\%$, $0.1\%$, $1.95\%$ and $10.4\%$ on average at each noise level, showing that combining the two losses is a clear advantage for both tasks, while being able to use a single model for both A/V ASR and ASD.

\vspace{-4mm}
\section{Conclusions}
\vspace{-4mm}
\label{sec:conclusions}
We introduced a multi-task training setup for an Audio-Visual model that is capable of simultaneously performing automatic speech recognition and active speaker detection. The proposed architecture is particularly interesting as it provides a diarization signal for on-screen speakers, without requiring an explicit diarization model. We significantly improved on the active speaker detection classification accuracy without degrading the ASR performance. Our experiments show that a joint loss increases the accuracy on the ASD task by around $26\%$ when compared to the baseline state-of-the-art multi-person A/V ASR model~\cite{Braga21}, while actually increasing the ASR performance of the same model up to around $10\%$ on noisy scenarios.

\bibliographystyle{IEEEbib}
\bibliography{mybib}

\end{document}

%% file: math_commands.tex
\usepackage{amsmath,amsfonts,bm}

\def\eqref#1{equation~\ref{#1}}
\def\1{\bm{1}}

\DeclareMathAlphabet{\mathsfit}{\encodingdefault}{\sfdefault}{m}{sl}
\SetMathAlphabet{\mathsfit}{bold}{\encodingdefault}{\sfdefault}{bx}{n}
\newcommand{\tens}[1]{\bm{\mathsfit{#1}}}
\def\tA{{\tens{A}}}

\def\tF{{\tens{F}}}

\def\tI{{\tens{I}}}

\def\tQ{{\tens{Q}}}

\def\tS{{\tens{S}}}

\def\tV{{\tens{V}}}
\def\tW{{\tens{W}}}

\newcommand{\R}{\mathbb{R}}

\DeclareMathOperator*{\argmax}{arg\,max}